# Ligand Rebinding: Self-consistent Mean-field Theory and Numerical Simulations Applied to SPR Studies

**Manoj Gopalakrishnan¶[1], Kimberly Forsten-Williams§*, Theresa R. Cassino§, Luz Padro§, Thomas E. Ryan† and Uwe C. Täuber¶**

¶ Department of Physics, Virginia Polytechnic Institute and State University, Blacksburg, VA 24061-0435, USA.
§ Department of Chemical Engineering and Virginia Tech - Wake Forest University School of Biomedical Engineering and Sciences, Virginia Polytechnic Institute and State University, Blacksburg, VA 24061-0211, USA.
† Reichert, Inc., 3374 Walden Avenue, Depew, NY 14043, USA.
* Corresponding Author
        Kimberly Forsten-Williams
        Department of Chemical Engineering and Virginia Tech - Wake Forest University
            School of Biomedical Engineering and Sciences
        Virginia Polytechnic Institute and State University
        Blacksburg, VA 24061-0211, USA.
        540-231-4851 (tel), 540-231-5022 (fax), kfw@vt.edu

**Abstract**
Rebinding of dissociated ligands from cell surface proteins can confound quantitative measurements of dissociation rates important for characterizing the affinity of binding interactions. This can be true also for *in vitro* techniques such as surface plasmon resonance (SPR). We present experimental results using SPR for the interaction of insulin-like growth factor-I (IGF-I) with one of its binding proteins, IGF binding protein-3 (IGFBP-3), and show that the dissociation, even with the addition of soluble heparin in the dissociation phase, does not exhibit the expected exponential decay characteristic of a 1:1 binding reaction. We thus consider the effect of (multiple) rebinding events and, within a self-consistent mean-field approximation, we derive the complete mathematical form for the fraction of bound ligands as a function of time. We show that, except for very low association rate and surface coverage, this function is non-exponential at all times, indicating that multiple rebinding events strongly influence dissociation even at early times. We compare the mean-field results with numerical simulations and find good agreement, although deviations are measurable in certain cases. Our analysis of the IGF-I-IGFBP-3 data indicates that rebinding is prominent for this system and that the theoretical predictions fit the experimental data well. Our results provide a means for analyzing SPR biosensor data where rebinding is problematic and a methodology to do so is presented.

---

[1] Present Address: Max Planck Institut für Physik komplexer Systeme, Nöthnitzer Straβe 38, 01187 Dresden, Germany.

## (i)    Introduction

Signal transduction via transmembrane receptor proteins is initiated by extracellular binding with specific proteins known as growth factors. These interactions tend to be of high affinity and, in many systems, are regulated by binding proteins present in the extracellular environment. Insulin-like growth factor-I (IGF-I) constitutes one prominent example of such a growth factor. Cell signaling is transmitted by direct interaction with the IGF-I receptor but this binding can be impacted by solution and cell-associated IGF binding proteins (IGFBPs), of which there are at least six. Quantification of the interactions of IGF-I with IGFBPs is critical if one is to understand how changes in expression and secretion will impact IGF-I signaling. Surface plasmon resonance (SPR) is one technique amenable to such measurements. SPR is an optical sensor technique that has the advantage of being able to take real-time measurements using low concentrations of unlabeled biologicals [reviewed in Cooper 2003].

Quantification of IGF-I interactions with both cell surface receptors and IGFBPs using SPR has been performed as a means of evaluating and predicting the competition between these molecules for IGF-I. Studies have used immobilized IGF-I [Wong et al. 1999; Dubaquie and Lowman 1999; Galanis et al. 2001; Fong et al. 2002; Vorwerk et al. 2002], IGF-I receptor [Jansson et al. 1997], or IGFBPs [Heding et al. 1996; Jansson et al. 1997; Marinaro et al. 1999; Fong et al. 2002; Vorwerk et al. 2002] using amine chemistry to link the proteins to a carboxymethyl dextran (CMD) layer on the SPR chip. Deviations from a single reversible binding model have been noted and attributed primarily to non-uniform coupling of the ligand to the gel. Fong et al. (2002) compared kinetic parameters for IGF-I and IGFBP-1 using both a CMD and a self assembled monolayer (SAM) chip and saw significant differences in derived binding affinities that they attributed to possible steric hindrance effects and transport issues. Vorwerk et al. (2002) used a CMD chip with coupled IGFBP-3 and measured values that differed from previous work [Heding et al. 1996; Wong et al. 1999; Galanis et al. 2001; Fong et al. 2002] that they attributed to the use of increased flow rate to assist in combating mass transport and rebinding effects. However, regardless of flow rate, fitting of dissociation data for this system has been problematic.

A phenomenon of particular interest in the quantification of ligand interactions is rebinding: a ligand, following dissociation from a bound protein on the surface, may diffuse in the extracellular fluid environment for some time and may be reabsorbed later at one of the free binding sites. Rebinding is believed to be an important mechanism in producing cellular response, especially with dilute ligand concentrations, by assisting receptor proteins to stay in the active state for longer periods of time. Rebinding also may promote co-operative behavior among clustered receptors by reducing the overall ligand dissociation, a phenomenon observed recently in experiments addressing the role of clustering in lipid rafts [Chu et al. 2004].

From a more general perspective, a quantitative characterization of the effects of rebinding is important in experiments like SPR, when dissociation rates of growth factors (or other ligands) are measured. In such a situation, it would be ideal to eliminate rebinding altogether since it interferes with the measurement of intrinsic dissociation and might lead to imprecise and significantly reduced dissociation rates [Nieba et al.1995]. Low surface coverage and higher flow rates are techniques used to counteract mass transport limitations [Schuck 1997]. Further, inclusion of specific proteins or molecules that may be used to bind to the released ligands or un-occupied binding sites and thus make them unavailable for rebinding is another technique targeted specifically at the rebinding problem. This technique has been used successfully for measuring the interaction of the SH2 domain of lck with a phosphotyrosine peptide [de Mol et al. 2000]. However, in the absence of quantitative information on the affinity of these agents for binding to either the ligand or the receptor, it is difficult to estimate the general reliability of these methods. An alternative is to understand how much rebinding might alter the dissociation of ligands in a given environment, and use this information to estimate the intrinsic rate of dissociation.

Rebinding of ligands to cell surface receptors has been extensively studied before in the context of isolated cells in a solution of ligands. Berg and Purcell [Berg and Purcell 1977] showed that the association rate of ligands in this case initially increases with the receptor number N (per cell), and approaches a finite value in the limit of large N (corresponding to a cell surface completely covered by receptors). Similarly, the effective dissociation rate of ligands from cell surface receptors was shown to be dependent on N, and is, in general, smaller than the dissociation rate from isolated receptors in solution [De Lisi and Wiegel 1981; Shoup and Szabo 1982; Goldstein et al. 1989; Zwanzig 1990; Goldstein et al. 1999]. This non-trivial effect is attributed to increased rebinding of ligands in the case of cell surface receptors: a dissociated ligand is likely to return to the cell surface several times over a small interval of time before diffusing far away from it. This causes a reduction in the effective dissociation rate, which increases in significance as the receptor density is increased.

The role of rebinding is further enhanced when the effective dimensionality of the interaction between ligands and receptors is reduced. For instance, consider a single layer of cells in a tissue or in a petri dish. The ligands diffusing in the local cell vicinity will bind to sites in this cell layer, which is effectively a two-dimensional plane over sufficiently small (but non-microscopic) length scales. This feature is particularly relevant in experimental methods such as SPR, where the binding proteins (receptors) are attached to a planar surface. The rebinding phenomena in this (2+1)-dimensional geometry must be expected to be qualitatively different from the full three-dimensional situation studied by previous authors, since the return-to-the-origin characteristics of a random walk are strongly dimension-dependent [Feller 1966].

Ligand rebinding in the case of receptors on a planar surface has been addressed in a few previous studies [reviewed in Goldstein et al. 1999]. For example, competition between convective and diffusive aspects of transport in the Biacore biosensor was studied by Edwards et al. [Edwards et al. 1999] in the limit of large flow velocity. Wolfsy and Goldstein [Wolfsy and Goldstein 2002] studied the effective rate coefficients in a Biacore experiment where the receptors are attached to polymers immobilized on the sensor surface. A rigorous mathematical study of ligand rebinding to receptors attached on a planar surface (in the absence of flow and solution receptors) was presented by Lagerholm and Thompson [Lagerholm and Thompson 1998]. In this work, coupled partial differential equations were used to study the time evolution of the probability of rebinding with appropriate boundary conditions at the surface. Although various quantities such as the rebinding probability of a released molecule could be calculated within this formalism, these expressions could not be directly compared to existing experimental results, where typically only the bound fraction of ligands is measured.

In this paper, we present an alternative formalism to study ligand dissociation from receptors attached to a planar surface in the limit of vanishing flow velocity, i.e., fully diffusion-limited transport. In contrast to most previous approaches, we describe the rebinding dynamics in terms of Brownian trajectories of individual ligand molecules dissociating from and re-attaching to the surface, with possibly multiple visits to the surface in between. Within a mean-field approximation, this approach yields a self-consistent integral equation for the fraction of bound receptors as a function of time, whose general solution is a slowly decaying non-exponential function. Monte Carlo simulations confirm the non-exponential nature of the decay. Experimental results of SPR experiments designed to measure dissociation of IGFBP-3 from IGF-I are presented, which are performed (i) in both the presence and absence of flow, (ii) with and without addition of soluble heparin in the dissociation phase to bind released IGFBP-3 in solution, and (iii) with varying surface coverage of IGF-I. The theoretical dissociation function is checked against the experimental curves, both in the presence and absence of exogenous heparin (that binds to IGFBP-3 but not IGF-I [Forsten et al. 2001]). The agreement is found to be very good, in the presence and absence of flow, up to time scales ~ 200-300 s at which time other features, perhaps the finite height of the experimental system, appear to become significant and further slowing down of the dissociation is evident. Our results therefore indicate that a proper assessment of rebinding effects is crucial in the analysis of SPR dissociation data, which might otherwise lead to erroneous estimation of rate constants.

This paper is divided into the following sections. In (ii), we describe the SPR experimental setup and results. In (iii), our self-consistent mean-field theory is presented in detail and the mathematical form for the full dissociation curve is obtained in that framework. We then analyze the data by means of the mean-field function. Sec. (iv) is concerned with the simulation model and the numerical results. We summarize this work and our findings in (v).

## (ii) Experimental

Growth factor signaling is regulated by both association and dissociation with cell signaling receptors with both rates impacting the persistence of the interaction [Lauffenburger and Lindermann 1993]. Measurement of these kinetic rates *in vivo* is difficult and confounded by potential alternative binding partners leading experimentalists to use techniques such as SPR to measure the isolated interactions. The ratio of the dissociation rate and association rate constants for a 1:1 binding reaction at equilibrium is referred to as the equilibrium dissociation constant ($K_D$), and it can be used in conjunction with either the dissociation or association rate to determine the other rate constant. Measurement of $K_D$, however, can be time-consuming using SPR and independent measurement of the rate constants would be preferable. This is the approach used in the experiments reported in this paper.

**2.1 Surface preparation**
The surfaces used for these studies were composed of a mixed self assembled monolayer (mSAM) on gold (500 nm) coated slides (EMF Corporation, Ithaca, N.Y.) prepared as previously described [Lahiri et al. 1999]. Briefly, the gold coated slides were immersed in a mixture of 0.2 mM carboxylic acid-terminated thiol and 1.8 mM tri(ethylene glycol)-terminated thiol (Toronto Research Chemicals, Toronto, Canada) for 12 hours. The surfaces were then rinsed with ethanol and dried under nitrogen. The resulting surface had free carboxyl groups for amine coupling and polyethylene glycol to minimize non-specific binding (Fig. 1).

**2.2 Activation and immobilization**
Activation of the surface was achieved using N-ethyl-N-(3-diethylaminopropyl) carbodiimide (EDC) and N-hydroxysuccinimide (NHS) chemistry. Immobilization was done both on-line and off-line. Briefly, off-line immobilization was initiated by washing the chip surface with 20 mM NaOH and rinsing with phosphate buffered saline with 0.005% Tween, pH 7.4 (PBST) (Sigma-Aldrich Corp., St. Louis, MO). A fresh solution of 0.2 M EDC (Pierce, Rockford, IL) and 5 mM NHS (Aldrich Chemical Co., Milwaukee, WI) was placed on the surface of the slide and allowed to react for 12 min at room temperature. The chip was then rinsed with 20 mM sodium acetate, pH 5.5. IGF-I (PeproTech, Inc., Rocky Hill, New Jersey) was then immobilized by placing 0.2 ml of 3.3 µM IGF-I in 20 mM sodium acetate solution onto the surface and incubated overnight in a container sealed under nitrogen at 4 °C. Following a wash with PBST, the slide was rinsed with 1M ethanolamine and then deactivated by surface exposure to 1M ethanolamine for 10 min at room temperature. The surface was then washed several times with PBST and dried with nitrogen prior to placing on the SPR unit.

On-line immobilization was performed in a similar fashion. Briefly, after placing the chip on the sensor surface, on-line immobilization was initiated by washing the chip surface with deionized water and then switching to PBST for ~5 min. until a stable baseline SPR signal was obtained. EDC/NHS solution (0.2 M EDC and 5 mM NHS in deionized water) was then injected into the system to activate the surface and allowed to react for 10 min. 20 mM sodium acetate buffer (pH 5.5) was then run over the sensor surface for ~ 5 min. until a stable baseline was obtained. IGF-I was then immobilized by running 3.3 µM IGF-I in 20 mM sodium acetate solution over the surface for a particular amount of time to obtain the amount of IGF-I desired on the surface. PBST was then run for 4 min. to wash the surface. Following the PBST wash, 1M ethanolamine was run for 10 min. to deactivate the surface and prevent covalent binding of other proteins to the slide. The surface was then washed with 20 mM HCl and 20 mM NaOH for 5 minutes each before switching to PBST for the binding experiments. The entire procedure was carried out at 25°C (controlled by the SPR instrument).

**2.3 Dissociation experiments**
Dissociation experiments were performed on a Reichert, Inc. SR 7000 Alpha instrument (Buffalo, New York) following either off-line immobilization of IGF-I and chip placement on the unit or on-line immobilization of IGF-I. PBST was run over the sensor surface and then followed by washes with 20mM HCl and 20 mM NaOH for 5 min each. The system was returned to PBST until a stable baseline was obtained (< 10 minutes) at a flow rate of ~0.8 ml/min. IGFBP-3 (Upstate Biotechnology, Lake Placid, NY) was pumped over the surface for 15 min to allow association. Following the association phase, PBST or PBST with heparin sodium from porcine intestinal mucosa (Celsus, Cincinnati, OH) was pumped over the surface to promote dissociation of the bound IGFBP-3. The surface was regenerated using 5 min washes of 2M NaCl in PBST, 20 mM HCl, and 20 mM NaOH. This procedure was repeated for each sample. Verification that the heparin sodium did not bind the IGF-I surface at the concentrations used for dissociation was performed.

**2.4 Experimental results**
Introduction of IGFBP-3 into the flow chamber over immobilized IGF-I led to the anticipated increase in binding characterized by a change in refractive index which is measured as pixels for the SR7000 Alpha unit, a similar unit to the RU commonly reported for the Biacore system (Fig. 1). The data was fit well by a 1:1 binding model with $R^2$ values of ~0.99 suggesting that heterogeneity of immobilized IGF-I was not a significant issue. However, the dissociation phase did not reflect the exponential decay one would expect for a 1:1 binding interaction and global fitting using CLAMP [Myszka and Morton, 1998] either with or without mass transport did not provide a good fit (data not shown). Similar dissociation phase data were collected whether the system was under high flow or not (Fig. 2) despite some differences in the kinetics of association (data not shown) suggesting that rebinding might be a more prominent issue in the deviation from expected results for the dissociation phase.

We have shown previously that IGFBP-3 and heparin interact strongly and negligible binding affinity exists between IGF-I and heparin, suggesting heparin might be a good rebinding inhibitor [Forsten et al. 2001, Goldstein 1989]. Inclusion of heparin in the association phase significantly reduced IGFBP-3 binding levels while no change in pixels was observed when heparin was introduced to the IGF-I flow cell in the absence of IGFBP-3 (Fig. 2B). We therefore investigated whether inclusion of heparin in the dissociation phase fluid would impact the dissociation rate. A significant reduction was observed (Fig. 3) and was repeatable for both multiple runs on the same and on different chips and in the presence and absence of flow. A heparin dose dependence effect was seen (data not shown). The reduction, however, still did not reflect exponential decay over the entire time regime. We emphasize that these experiments were all done with off-line coupling of IGF-I to obtain high coverage, and consequently good signal-to-noise, for our system.

The SAM is designed with only 10% of sites available for binding, although, depending on the radius of the protein, higher overall surface densities/coverage are possible [Lahiri et al. 1999]. We, therefore, used on-line coupling to reduce coverage to see if that might impact dissociation. Ligand loading has been shown previously with CMD surfaces to impact interaction kinetic measurements [Edwards et al. 1997]. A reduction in IGF-I surface coverage did result in somewhat faster dissociation particularly in the presence of high heparin concentrations but exponential decay was still not observed (Fig. 4). Normalized association curves, moreover, were not significantly different for the reduced coverage chips (data not shown).

To summarize, the SPR experiments with IGF-I and IGFBP-3 provide strong evidence that rebinding of ligands can significantly affect the dissociation data. Rebinding was found to be important even in the presence of flow conditions as well as high heparin concentrations. In the following section, we present a systematic theoretical treatment of this system which accounts for the observed non-exponential decay of the dissociation curve.

## (iii) Theory

**3.1 Self-consistent mean-field theory of ligand rebinding**
In this section, we outline the mathematical theory of ligand dissociation and the consequent (multiple) rebinding to the binding sites on the surface. We attempt to simulate the SPR experimental set-up (Fig. 1) and model the immobilized surface proteins (receptors or binding sites) as being homogeneously distributed on a two-dimensional surface (Fig. 5). We employ a self-consistent mean-field approximation that takes into account the full binding-dissociation-rebinding history of a given ligand, yet ignores the details of the spatial organization of the receptors. This means that a ligand molecule infinitesimally close to the surface binds to it with a space-independent probability, which depends only on the mean fractional surface coverage and the association rate. By construction, this approximation works better when the fractional coverage of receptors is low, so that the probability of occurrence of regions with surface coverage significantly larger than the mean value is small. Our numerical simulations show that the approximation works reasonably well up to about 10% of fractional coverage, but the deviation is more significant at 50% coverage (ref. Fig. 6 below).

To simplify the analysis, we neglect the flow conditions in the system, i.e., our model is strictly valid only in the limit where the flow velocity vanishes. However, it is possible that this is not a serious limitation since, on account of the viscous drag, the flow velocity decreases towards the substrate, and vanishes altogether at the surface. This presumably would lead to the formation of a diffusion-limited zone close to the sensor surface, where mass transport is controlled by diffusion rather than convection. Our theory should describe well such a system, and we further note that the experimental results did not show any significant difference in dissociation between the flow and no-flow situations, further justifying this approximation.

We define $p(t)$ to be the fraction of binding sites which are bound to ligands at time t, so that p(0) is the fraction bound immediately following association. The dissociation and association rates are denoted $k_-$ and $k_+$ respectively. The most general equation describing the time evolution of $p(t)$ is then

$$\frac{dp(t)}{dt} = -k_- p(t) + k_+ \rho(\varDelta,t)[1-p(t)] \ , \tag{1}$$

where $\rho(z,t)$ denotes the ligand concentration (throughout this paper, we define concentration as number of molecules per unit volume, rather than the more conventional number of moles per unit volume noting that a simple conversion factor can be used to interconvert between the two units) as a function of the perpendicular distance z from the surface and time t. The length scale $\varDelta$ is a microscopic length scale (defined more precisely in the following paragraph and section 3.3), and $\rho(\varDelta,t)$ is the ligand concentration within a thin slab of thickness $\varDelta$ above the surface. The first term in Eq.1 corresponds to dissociation and the second term represents the rebinding events.

We now propose a lattice formulation of the problem, whereby the ligand diffusion is modeled as a discrete random walk in three-dimensional space. For simplicity, we ignore the detailed three-dimensional structure of the receptors (and ligands). Instead, the two-dimensional substrate surface is modeled as a (square) lattice of randomly mixed potential binding sites (depending on occupancy) and non-binding sites. The lattice spacing for the full three-dimensional lattice is defined to be $\varDelta$, which may be interpreted as the typical distance traversed by a ligand molecule in solution without appreciable change in its direction (defined more precisely in Sec.3.3). The fraction of binding sites in the substrate lattice is denoted by $\theta$ and is proportional to the surface coverage of the receptors. However, since the binding process is not necessarily diffusion-limited, it would be more accurate to regard $\theta$ as an effective parameter which is a function of the association rate $k_+$. The relation between the two quantities ($\theta$ and $k_+$) is crucial to our analysis, and will be discussed in detail in Appendix A. A glossary of the important parameters and symbols used in this paper is provided in Table 1.

Let us define the rebinding rate $\gamma(t) = k_+ \rho(\varDelta,t)$, which we will now compute within the lattice model. The basic stochastic event contributing to the rate $\gamma(t)$ is the dissociation of a ligand at a certain bound receptor during a given time interval $[\tau;\tau+d\tau]$, where $\tau < t$, and its subsequent adsorption at the reference site at time t. In the lattice formulation, this process is a random walk starting at the point $(x,y,\varDelta)$ and ending at the origin, with possibly multiple visits to the surface in between. We have also assumed here that the vertical separation between a ligand and the substrate surface immediately following dissociation is of the order of $\varDelta$.

Within the mean-field approximation we employ here, the spatial fluctuations in receptor density are ignored. The rebinding probability in this case may be expressed as

$$\gamma(t) = k_- \int_0^t d\tau\, p(\tau) C_\theta(\Delta; t-\tau)\ , \tag{2}$$

where $C_\theta(\Delta; T)$ is the probability that a diffusing particle starting at the point $z = \Delta$ at time $t = 0$ is adsorbed at $z = 0$ at $t = T$ (see Appendix B for details). Combined with Eq.1, we thus have a self-consistent equation for $p(t)$. From Eq.2, we also note that the ligand concentration close to the surface is given by $\rho(\Delta, t) = K_D \int_0^t d\tau\, p(\tau) C_\theta(t-\tau)$ where $K_D = k_-/k_+$ is the standard equilibrium dissociation constant, with units of M. A more complete discussion of the density profile of the ligands in solution is presented in Appendix C.

A note on the boundary conditions of the problem is required at this point. A potential binding site would become non-binding after adsorbing a ligand, and would become a binding site again after releasing this ligand. This means that, within the mean-field formulation, the probability of adsorption and reflection are time-dependent: the plane $z=0$ absorbs the particle with probability $\theta[1-p(t)]$ and reflects it with probability $1-\theta+\theta p(t)$. Our problem is similar to that addressed previously by Agmon [Agmon 1984]): however, there are important differences. Agmon studied the 'survival probability' of ligands (in our language) rather than the occupancy of receptors. Further, a trial solution linear in the reaction probability was used, which could only be solved for specific cases. Our formulation is more general and not restricted to specific cases.

In order to further simplify the setup, we assume that the initial bound fraction $p(0) \ll 1$, (see, however, the discussion at the end of Appendix D) so that the absorption and reflection probabilities are effectively time-independent. In this limit, rebinding of ligands is effectively reduced to a well-defined one-dimensional random walk problem. Note that, in contrast to previous approaches to the rebinding problem, we do not need to take into account the density profile of the ligands in solution (see Appendix C, however). Rather, by describing the dynamics in terms of trajectories of individual Brownian particles, we arrive at an elegant non-Markovian effective equation for the dissociation curve itself with a minimum of assumptions. In particular, the need for non-trivial boundary conditions at the surface is eliminated.

Equations 1 and 2 combined are formally solved using Laplace transforms. Let us define the Laplace-transformed variables $\tilde{p}(s) = \int_0^\infty p(t) e^{-st} dt$ and $\tilde{C}_\theta(s) = \int_0^\infty C_\theta(\Delta; t) e^{-st} dt$. In terms of these variables, Eq. 1, after substituting Eq. 2, becomes

$$s\tilde{p}(s) - p(0) = -k_- \tilde{p}(s)\left[1 - \tilde{C}_\theta(s)\right]\ , \tag{3}$$

which leads to

$$\tilde{p}(s) = \frac{p(0)}{s + k_-\left[1 - \tilde{C}_\theta(s)\right]} \quad . \tag{4}$$

The next step in our calculation is to compute $\tilde{C}_\theta(s)$, which we accomplish as follows: Let us consider a one-dimensional random walk on the semi-infinite line $0 < z < \infty$ and define $q(\Delta,t)$ as the probability that a random walker, starting at position $z = \Delta$ at time t = 0, will visit the origin again for the first time at instant t. The probability of absorption of the random walker at the origin at time is $C_\theta(\Delta,t)$, which may be expressed in terms of $q(\Delta,t)$ via the following self-consistent equation:

$$C_\theta(\Delta,t) = \theta q(\Delta,t) + (1-\theta)\int_0^t d\tau\, q(\Delta,\tau)C_\theta(\Delta,t-\tau) \quad . \tag{5}$$

The first term in this expression gives the probability that the ligand will be re-absorbed at its first attempt to make contact with the surface. The second term is the sum of the probabilities of all the other events where the ligand is reflected at the first attempt (either because the surface site is non-binding or is already bound to another ligand), which may happen at any intermediate time $\tau$, but is adsorbed at one of the subsequent attempts at time t. Using Laplace transforms as before, this expression becomes $\tilde{C}_\theta(s) = \theta\tilde{q}(s) + (1-\theta)\tilde{q}(s)\tilde{C}_\theta(s)$, from which we infer

$$\tilde{C}_\theta(s) = \frac{\theta\tilde{q}(s)}{1-(1-\theta)\tilde{q}(s)} \quad . \tag{6}$$

The first passage probability for a random walker in one dimension is a well-studied problem with known result, namely: $q(z,t) = \frac{z}{t}\frac{1}{\sqrt{4\pi D t}}e^{-\frac{z^2}{4Dt}}$ [Feller 1966] for any $z > 0$ where $D$ represents the diffusion coefficient for the three-dimensional random walk.. Upon performing the Laplace transform of this expression, we find that

$$\tilde{q}(s) = e^{-2\sqrt{\delta s}} \quad , \tag{7}$$

where we have introduced the quantity

$$\delta = \frac{\Delta^2}{4D} \quad , \tag{8}$$

which constitutes a microscopic time scale in the problem, which is computed in Sec.3.3. After substituting Eq. 7 into 6, we arrive at

$$\tilde{C}_\theta(s) = \frac{\theta e^{-2\sqrt{\delta s}}}{1-(1-\theta)e^{-2\sqrt{\delta s}}} \quad . \tag{9}$$

Note that this quantity vanishes in the limit $\theta \to 0$ since adsorption becomes rare in this case. This means that, in principle, rebinding can be effectively prevented and exponential dissociation recovered at sufficiently small times in the case of low surface coverage of receptors. This case is discussed in detail in Appendix D. For the rest of this section, we assume that $\theta$ is sufficiently large so that rebinding is significant.

With Eq. 9, we have, in principle, solved the rebinding problem under the mean-field approximation. However, the resulting general expression obtained after substituting Eq. 9 into Eq. 4 is too complicated to invert to find the rebinding probability. Fortunately, without much loss of generality, we can assume that the microscopic time scale $\delta$ is sufficiently small (in comparison with $k_-^{-1}$) so that the approximation $e^{-2\sqrt{\delta s}} \approx 1 - 2\sqrt{\delta s}$ can be used. With this simplification, we find that $1 - \widetilde{C}_\theta(s) \approx (2/\theta)\sqrt{\delta s}$ (when s is sufficiently small, see Appendix D ). After substituting this expression into Eq. 4, we arrive at the final result

$$\widetilde{p}(s) = \frac{p(0)}{s + \frac{2k_-}{\theta}\sqrt{\delta s}} \quad . \tag{10}$$

We note from this expression that the readsorption events have strongly modified the dissociation curve. In the absence of rebinding, this expression would simply read $\widetilde{p}(s) = p(0)/(s + k_-)$, which is just the Laplace transform of an exponential decay curve, $p(t) = p(0)e^{-k_- t}$. In other words, the effect of rebinding is not simply a reduction of the effective dissociation rate, but rather leads to a non-exponential temporal decay of the bound fraction. This is explicitly seen after inverting Eq. 10, which yields

$$p(t) = p(0)e^{ct}\,erfc(\sqrt{ct}), \quad \text{where} \quad c = \frac{4k_-^2 \delta}{\theta^2} \quad \text{and} \quad erfc(z) = \frac{2}{\sqrt{\pi}}\int_z^\infty e^{-x^2}dx \quad . \tag{11}$$

This final expression is thus characterized by a single effective time scale $1/c$, which is proportional to the inverse of the square of the dissociation rate. We also note that dimensional consistency is ensured by the introduction of $\delta$, which is essentially the smallest time scale over which a rebinding event takes place. The rebinding processes, therefore, alter the temporal behavior of the dissociation curve in a fundamental way.

Within the limitations of the mean-field approximation and the assumptions used so far, Eq. 11 constitutes a complete solution of the rebinding problem, over sufficiently large time scales. Over very small time scales, the solution displays exponential decay (ref. Eq. D2), whereas over intermediate time scales, $t_e \ll t \ll 1/c$ the decay is an expanded stretched exponential, as seen by applying the small-argument expansion of the complementary error function [Abramowitz and Stegun 1970]: $erfc(z) \cong 1 - (2/\sqrt{\pi})z + o(z^2))$, leading to $p(t) \approx p(0)[1 - (4k_-/\theta\sqrt{\pi})\sqrt{\delta t} + ...]$. The combination of the exponential and stretched exponential decay at early times suggests that naïvely fitting the initial part of the decay curve to an exponential is likely to significantly under-estimate the dissociation rate. For very late times (t>>1/c), the decay becomes a power law, i.e., $p(t) \sim 1/\sqrt{t}$. This regime is indeed observed in numerical simulations when a very low value of the coverage $\theta$ is used (c.f. Figs. 8 and 9 below).

We would also like to add a note on finite-size effects here, keeping in mind that our goal is a direct comparison with experimental results (to be discussed in Sec. iv). The experimental system has a finite 'height', so that dissociated ligands which wander too far from the adsorption surface are eventually reflected back under no-flow conditions. This effect of the boundary will be seen in the dissociation curve after a certain crossover time scale, which we estimate as typically $\tau_H \sim H^2/D$, where H is the sample chamber height. The presence of the boundary thus leads to additional rebinding events (as compared with the idealized case of infinite H studied so far), and slows down dissociation even more relative to our mean-field prediction over times $t >> \tau_H$. This deviation from the infinite-height mean-field prediction is indeed observed in the experiments (Sec. iv). An extension of the present study that takes the finite height of the experimental system into account is currently in progress.

### 3.2 Lattice model of ligand-receptor binding

We next describe a discrete lattice model amenable for simulating the rebinding problem. These simulations are done in order to check the validity of the self-consistent mean-field assumption employed in the analytical treatment above. As mentioned in the introduction to the last section, we imagine the SPR slide surface as a two-dimensional square lattice of dimensions $L \times L$. A fraction $\theta$ of the lattice sites are occupied by receptor proteins, i.e., they serve as potential binding sites for the ligand. The remainder of the sites are non-binding: the ligands, upon contact with one of these, will be reflected back to the solution. The ligands themselves are modeled as Brownian particles (random walkers) diffusing in the semi-infinite three-dimensional space of which the cell surface forms one (partially absorbing) boundary. Periodic boundary conditions are imposed on all four borders of the two-dimensional lattice so that a ligand that exits at one boundary will reenter the system from the opposite one. The direction perpendicular to the plane of the lattice shall be referred to as the z-axis, and the surface itself is positioned at $z = 0$ (c.f. Fig. 5). Ligand diffusion in the z-direction is not bounded. As the receptor proteins are covalently attached to the surface, we treat them as static in this study ignoring any position fluctuations or movements. The lattice dimension is fixed at L=100 for all the simulations reported in this paper.

At the beginning of the dynamics, a fraction *p(0)* of all the binding sites are bound to a single ligand each, i.e., the total number of ligands in the system is $N = L^2 \theta p(0)$, and is conserved throughout the simulation. There are three main dynamical processes in the simulation:

(i) Dissociation of a ligand from a bound receptor: this process takes place with probability $\tilde{\beta}$ per time step. This move updates the position of the ligand from z = 0 to z

= 2. (We set z = 2 instead of z = 1 in order to prevent immediate rebinding to the same receptor.)

(ii) Diffusion of a released ligand in solution: a free ligand moves a distance equal to one lattice spacing in one of the six possible directions (i.e., to nearest-neighbor sites in the cubic lattice) with probability $\tilde{D}=1/6$ per time step.

(iii) Readsorption of free ligand to a free receptor: A free ligand at z = 1 (or correspondingly, z = $\Delta$ in the continuum theory of last section) is absorbed by a free receptor below it, if there is one present at that site, with probability unity, i.e., the ligand-receptor binding is assumed to be purely diffusion-limited: the binding reaction always occurs when possible.

### 3.3 Parameters in the lattice model

In order to establish a close connection between the lattice model in our simulations and the underlying experimental system, it is necessary to put our choice of parameters on a firm footing. We begin with the microscopic length scale $\Delta$, which we define as the distance moved by a ligand following dissociation, before a significant change in its direction of motion takes place. The time taken by the ligand to travel this distance is then simply equal to $2\delta = \Delta^2/2D$. For a Brownian particle of mass $m$ moving in a fluid of viscosity $\eta$, the velocity correlations decay exponentially fast: $\langle \vec{v}(0)\cdot\vec{v}(t)\rangle \propto e^{-t/\tau}$, with

$$\tau = m/6\pi a\eta = Dm/6kT , \qquad (12)$$

where we have used the Stokes-Einstein formula $D = k_B T/\pi\eta a$ to eliminate the 'radius' of the ligand molecule $a$ in favor of the diffusion coefficient D. Over a time scale $\sim 10\tau$, the directional correlation in the motion of the ligand is lost. Following our previous argument, this time scale is simply $2\delta$. Such an operational definition would yield $\delta \approx 5\tau$ (Combined with Eq.8, this expression also defines the length scale $\Delta$). The mass of an IGFBP-3 molecule is about 47kDa. The diffusion coefficient can be estimated from the finite-size effect observed in the experimental dissociation curve and here we simply quote the value obtained: $D \approx 1.5\times 10^{-9} m^2 s^{-1}$ (Eq. 14 - see Sec iv. for calculation). Since the experiments are done at room temperature, T ~ 300K. After substitution of these values in Eq.12, we find

$$\delta \approx 10^{-11} s . \qquad (13)$$

In the simulations, we also chose a dissociation rate per unit time step, $\tilde{\beta} \approx k_{\_}\delta t$, where $\delta t = \Delta^2/6D = 2\delta/3$ is the diffusion time scale. 'Typical' dissociation rates reported in the literature are quite small on the scale of $1/\delta$ - for example, Vorwerk et al. estimated $k_{\_}$ for IGF-I-IGFBP-3 using SPR to be ~0.01 min$^{-1}$ [Vorwerk et al. 2002]: this means that we may expect $\tilde{\beta} \ll 1$ generally, since $\delta$ is a microscopic time scale. In the simulations reported here, we have chosen $\tilde{\beta} = 10^{-5}$ to limit computational time (the value of $\tilde{\beta}$ matching experimental conditions is likely to be smaller by several orders of magnitude). For the simulations, starting from a randomly distributed set of receptors, the dynamics is carried out up to $10^7$ Monte Carlo (MC) time steps (i.e., up to 100 times $1/\tilde{\beta}$). The bound fraction *p(t)* is measured every 100 MC steps. The resulting dissociation curve is then averaged over 20 different starting configurations.

### 3.4 Simulation results
In Fig. 6, we show the (normalized) dissociation curve as obtained from the Monte Carlo simulations, for two values of $\theta$, 0.1 and 0.5 respectively, plotted against the 'scaled time' $T = \tilde{\beta} t$ where t is the number of MC steps. The fraction of surface proteins initially bound to diffusible ligand was fixed at p(0) = 0.25. We find that, everything else being the same, larger $\theta$ results in stronger rebinding and hence slower dissociation. Since time is measured in units of $k_-^{-1}$, the effective fitting parameter becomes $c = 4k_-\delta/\theta^2 = 6\tilde{\beta}/\theta^2$. For $\theta$ = 0.1 and 0.5, respectively, the theoretical fitting parameters are thus 0.006 and 0.00024. The measured values found using the best fits to the simulation data are close, but somewhat larger than these theoretical values. This slight discrepancy could be due to two factors: (i) the mean-field calculation assumes that all the surface proteins are available for rebinding at any given time, whereas in the simulations only free receptors are available; (ii) a systematic deviation from the mean-field prediction might exist, since (especially for high surface protein densities), local density fluctuations are likely to become important in the rebinding.

In Fig. 7, we show how the dissociation curves behave in the case of an extremely small fraction of binding sites ($\theta = 0.01$) on the surface, when the dissociation rate is varied. In this situation, two regimes are observed. When the dissociation rate $\tilde{\beta}$ is small, then between two dissociation events, the ligand has enough time to span the surface for binding sites. The dissociation curve is thus dominated by rebinding, and a very slow decay in accordance with Eq. 11 is observed. Alternatively, when dissociation is very fast, rebinding is very inefficient in competing with dissociation because the number of binding sites available is very small. The dissociation curve is, in this case, closer to the pure exponential dissociation curve in the absence of rebinding (see the discussion in Appendix D).

In Figs. 8 and 9, the dissociation curves are depicted for two small values of the coverage fraction: $\theta = 0.005$ and 0.01, holding $\tilde{\beta}$ fixed at $10^{-5}$. The logarithmic plots in Fig.8 show a crossover to the power-law decay $p(t) \sim 1/\sqrt{t}$ discussed in the previous section. While we would expect that such a crossover should occur in our experimental system, this might be difficult to observe because finite-size effects could disrupt and mask the entry into this regime.

### (iv) Results - Fitting the experimental data to the mean-field result

Having found good agreement between theory and simulations, we next investigated if the theory would also fit the experimental data to provide rationale for the lack of agreement between dissociation data and a 1:1 fit. We observe that, except for very late times, all the SPR data sets for IGF-I:IGFBP-3 were fit well by the theoretical prediction given by Eq. 11, with the parameter c suitably tuned (Figs. 10 and 11). Our next step was to determine how this could be used to analyze the experimental data. Namely, we wanted to estimate the intrinsic dissociation rate $k_-$ from the parameter c, and from Eq.11, this requires us to know the microscopic time scale $\delta$ (estimated in Sec. 3.3) and the effective surface coverage $\theta$. The latter is related to the association rate $k_+$ of ligands (when measured in exactly the same experimental setup), and could be estimated from the experimentally measured association rate and the actual surface coverage in the device (Appendix A). However, from Eq. A4 (Appendix A), this requires knowledge of the microscopic lattice length scale $\Delta = \sqrt{4D\delta}$ (Eq. 8): it is therefore crucial to have a reliable estimate of the diffusion coefficient, which we now attempt to obtain from the dissociation data themselves.

The late-time dissociation data (e.g., the zero heparin situation in Fig. 10) show a distinct flattening on account of the finite thickness of the SPR device chamber, which cause ligands which wander too far to bounce back to the system under no-flow conditions (whereas we had assumed this thickness to be infinite in the theoretical calculation). If the thickness of the sample is H, then the effects of this constraint on the perpendicular diffusion will start showing around a time scale $\tau_H \sim 2(H^2/2D)$, which represents the average time for a ligand molecule to diffuse to the boundary of the system *and* return to the surface after reflection. We may obtain an estimate of the diffusion coefficient, therefore, by determining $\tau_H$ from the data, as the first instant when a significant deviation of the experimental curve from the theoretical fit is seen. For the Reichert apparatus used in our experiments, the chamber height (thickness) was H=0.19mm. An estimate $\tau_H$ from Fig. 7 is $\tau_H \approx 230 s$. From these numbers, we estimate

$$D \approx 1.5 \times 10^{-9} m^2 s^{-1}. \tag{14}$$

A better way to estimate D is to determine the theoretical dissociation curve in the case of finite H and in the presence of the flow conditions, and use it to fit the experimental data. This will be carried out in a future work.

We now attempt to estimate the intrinsic dissociation rate $k_-$ from our curve-fitting analysis, using the fit value c = $1.9 \times 10^{-5}$ s$^{-1}$ for the heparin-free case (Fig. 10). After combining Eqs. 8, 12 and Eq. A3 from Appendix A, we see that the parameter c is a function of the equilibrium dissociation constant $K_D = k_-/k_+$:

$$c = 16D^3 \left(\frac{\delta K_D}{\theta_s}\right)^2 . \tag{15}$$

From Eq.12 and Eq.13, we then obtain $K_D \approx 3.11 \times 10^{-7} M$. The experimental value of the association rate measured with a similar sensor chip was $k_+ \approx 1.1 \times 10^7 M^{-1} min^{-1} = 3.05 \times 10^{-22} m^3 s^{-1}$, [Cassino 2002] (using the conversion $M^{-1} = N_A^{-1} \times 10^{-3} m^3$, where $N_A$ is the Avogadro number) with a fractional coverage $\theta_s \approx 0.1$. Substitution of these values results in the estimate $k_- \approx 3.43 min^{-1}$. This number is two orders of magnitude larger than a previous estimate [~0.01 $min^{-1}$, Vorwerk et al. 2002], which was obtained without taking rebinding events explicitly into account. Clearly, a better characterization of the experimental system would be necessary to obtain a more reliable quantitative estimate of the dissociation rate. Even so, our analysis shows that simple curve-fitting to an exponential decay might significantly under-estimate the dissociation rate.

Our detailed analysis suggests that, rebinding, even with small coverage fractions (which for our apparatus we estimate in the range of 10%), may significantly affect the dissociation process and its inclusion is needed in the interpretation of dissociation data. Not unexpectedly, the coverage fraction turns out to be an important parameter in this problem, and likely controls the difference between exponential and non-exponential behavior in dissociation. This crossover is more quantitatively characterized in Appendix D, where we also discuss the different parameter regimes where an exponential decay might be observed.

The addition of heparin in the buffer leads to faster dissociation (Fig. 11, also Table 2), and we observe a systematic increase in the fitting parameter c as the heparin level is increased. However, it is worth noting that an exponential decay is not recovered even with high heparin concentrations (1.8, 5.4, and 10.8 μM). This is all the more remarkable because the affinity of heparin for IGFBP-3 has been measured to be ~76 nM using affinity co-electrophoresis [Forsten et al. 2001]. In a well-mixed solution of heparin and IGFBP-3, the fraction of the free ligand in the steady state would be $p = 1/(1 + \rho/K_d)$, where $\rho$ is the heparin concentration. For $\rho = 1.8$ μM and 5.4 μM respectively, this fraction is only 0.04 and 0.01 (and similar values for the other heparin concentrations). If a steady state were indeed reached between heparin and the free ligand, we should see a corresponding change in the parameter $\theta$ since, presumably, only the ligand not bound to heparin will be available for rebinding (i.e., $\theta \to \theta_H = \theta p$). However, the change in $\theta$ as determined from the fitting parameter c is much less than this estimate noting that even at 10.8 μM heparin, about 15% of the ligand in solution are not bound to heparin and hence available for rebinding (Table 2).

### (v) Summary and Discussion

In this paper, we present an experimental, analytical, and computational study of the dissociation of ligands from a flat substrate. We have primarily focused on the role of potentially multiple rebinding of dissociated ligands, and how it affects the overall dissociation. The SPR experiments we modeled were performed with IGFBP-3 as the soluble ligand and IGF-I attached to a planar surface as the receptor. Porcine heparin was used to bind the dissociated IGFBP-3 in solution, and its effect on the dissociation at various concentrations was studied.

The dissociation of IGFBP-3 was non-exponential in all the SPR experiments performed (Fig. 3) despite using a planar geometry surface (Fig. 1) to reduce mass-transport limitations known to be problematic with SPR experiments [Schuck 1996; Schuck 1997]. It should be noted, however, that similar non-exponential dissociation results were found by us [Cassino 2002] and others [Wong et al. 1999; Dubaquie and Lowman 1999; Fong et al. 2002] using the more traditional carboxymethylated dextran slides with IGF-I or IGFBP-3 immobilized. The addition of heparin was observed to render the dissociation faster, presumably by binding the dissociated IGFBP-3 and preventing their rebinding to unbound IGF-I. However, in no experiment did we observe actual exponential decay. This was true even for heparin concentrations as high as 30 μM (> 1000 times the concentration of IGFBP-3 used in the association portion of the experiment) indicating that equilibrium was not reached between heparin and IGFBP-3 over the experimental time scales (data not shown). Simple fitting of an exponential to the dissociation data is, in general, not appropriate and a better tool is manifestly needed to determine quantitative values.

Our analysis, however, does not rule out the possibility of situations where an exponential fit to the dissociation curve might produce the right dissociation rate. As discussed in Appendix D, if the binding probability (i.e., the affinity of the receptor for the ligand) and/or surface coverage of receptors is small compared to the dissociation rate, the rebinding process affects dissociation only over very large times (ref. Eq. D2), and it may be possible to neglect it altogether. Alternatively, if the ligand has high affinity for an external binding agent, such as heparin in our system, then using this agent in sufficiently high concentrations could be successful in making rebinding insignificant as far as dissociation is concerned. Many examples are available in the literature where a simple model has been shown to fit well. For example, binding of interleukin-2 to a surface with immobilized IL-2 α-receptor on it was shown to fit well to a simple bimolecular model [Myszka 1999]. Alternatively, Schuck et al. [Schuck et al. 1998] use competitive dissociation to obtain an improved fit for binding of a specific Fab to immobilized whale neuraminidase.

The self-consistent mean-field theory presented in this paper provides a complete mathematical form of the dissociation curve in the presence of un-inhibited rebinding on a planar surface, in terms of a single effective parameter. This effective parameter has been shown to depend on the intrinsic dissociation rate, the effective surface coverage by receptor proteins (proportional to the association rate) and the ligand diffusion coefficient in solution. The history dependence of the dissociation process (rebinding of ligands depend on their dissociation from the surface at previous times) is rigorously taken into consideration by describing the ligand dynamics in terms of individual Brownian paths, rather than using the more conventional PDE approach [as in, e.g., Lagerholm and Thompson 1998]. As the formalism developed here yields the complete dissociation curve, we believe this to constitute a marked improvement over previous mathematical studies of rebinding, especially since our results could be directly compared with experiments. Our analysis also demarcates the different regimes in the full parameter space where rebinding is strong and weak and may be used in future SPR data analysis.

**Acknowledgements**

We thank D. Lubensky, R. Kree, T. Newman, H. J. Hilhorst, B. Schmittmann, and G. I. Menon for fruitful discussions. Financial support from the National Science Foundation [NSF-DMR 0089451 (MG), NSF-DMR 0308548 (UCT), NSF-9875626 (KFW), Graduate Fellowship (TRC)], National Institutes of Health [NIH-HL56200 (KFW)] and the Bank of America Jeffress Memorial Trust [Grant no. J-594 (UCT)] is gratefully acknowledged.

## Appendix A: Effective surface coverage and association rate

In this appendix, we show how the effective lattice coverage fraction $\theta$ may be related to the association rate $k_+$ in the continuum formulation. For this purpose, it is convenient to express the effective surface coverage in the form

$$\theta = \theta_s \theta_a , \quad (A1)$$

where $\theta_s$ is the actual fraction of binding sites in the lattice, and $\theta_a$ is the probability with which a ligand which comes infinitesimally close to a receptor by diffusion will bind to it before diffusion takes it away again.

Let us approximate the ligand motion as a discrete random walk with step size $\Delta$ (defined in text). Let $\mu_D$ be the rate with which the walk (projected onto the z-axis perpendicular to the surface) moves one lattice spacing. The unit time scale for one-dimensional diffusion is $\delta' = \Delta^2/2D$, and this move is made with probability ½. It follows that

$$\mu_D = \frac{1}{2\delta'} = \frac{D}{\Delta^2} . \quad (A2)$$

If the volume density (number per unit volume) of ligands infinitesimally close to the surface at time t is $\rho(\Delta,t)$, then the probability of finding a ligand in a volume element $v = \Delta^3$ is just $\mu_\rho = \rho(\Delta,t)\Delta^3$. A ligand at a 'height' $\Delta$ above a receptor will then bind to it at a rate

$$\gamma(t) = \mu_\rho \mu_D \theta_a = \rho(\Delta,t)\theta_a D\Delta . \quad (A3)$$

Note that, in the continuum formulation, the rate of binding is simply $\gamma(t) = k_+ \rho(\Delta,t)$ from Eq.1. Upon equating the two expressions, we arrive at the result

$$k_+ = \theta_a D\Delta , \quad (A4)$$

which defines $\theta_a$. This result may also be viewed as the one-dimensional analogue of the well-known result for diffusion-limited association rate on a spherical receptor of radius $b$ in three dimensions: $k_+(b) = 4\pi Db$ [Torney and McConnell 1983]. However, there is an important difference. Whereas the three-dimensional result is valid for a single isolated receptor molecule, Eq. A4 is valid only for a distribution of binding sites on a plane with a non-vanishing mean density. It would be, therefore, more correct to view Eq. A4 as an operational definition of the effective surface coverage for the lattice model.

## Appendix B: Calculation of the rebinding rate

Let us consider the bulk diffusion of a free ligand in three dimensions, starting at the point $(x,y,z)$ at time t = 0 and arriving at $(0,0,Z)$ at t = T. The probability density for this process will be denoted by $P(x,y,z;T)$; it is governed by the diffusion equation modified by a term to account for surface adsorption,

$$P(r,t+\delta t) - P(r,t) = \tilde{D}[\Sigma_{r'} P(r',t) - 6P(r,t)] - \theta \delta_{z,0} P(r,t) \, , \tag{B1}$$

where $\delta t$ denotes the microscopic diffusion time step, $\tilde{D} = D\delta t / \Delta^2$ is the effective diffusion coefficient for the underlying lattice (which we take to be cubic for simplicity) and $r = (x,y,z)$ represents the position of the particle in the three-dimensional space. The first term in Eq. B1 is simply the diffusion of the particle away from the surface, and the last term represents the adsorption at the surface that occurs with probability $\theta$. A schematic diagram of the set-up of our model is depicted in Fig. 5.

Since the space coordinates are clearly statistically independent here, the solution to Eq. A1 can be written in the form of a product, $P(x,y,z;T) = G_1(x;T)G_2(y;T)G_3(z,Z;T)$. Upon substitution in Eq. B1, we find, of course, that $G_1$ and $G_2$ satisfy the simple one-dimensional diffusion equation (without any adsorption), and only $G_3$ is modified by the adsorption term. The complete probability distribution may then be written as

$$P(x,y,z;T) = \frac{1}{4\pi DT} exp\left(-\frac{x^2+y^2}{4DT}\right) G_3(z,Z;T) \, . \tag{B2}$$

The rate of adsorption of the ligand at the surface (z=0) is simply $\tilde{P}(x,y;T) = D\partial_Z P|_{Z=0}$, and from Eq. B2, we infer that the derivative acts only on the function $G_3$. For a dissociated ligand, the initial position on the z-axis is $z=\Delta$. Hence, the absorption rate becomes

$$\tilde{P}(x,y;T) = \frac{1}{4\pi DT} exp\left(-\frac{x^2+y^2}{4DT}\right) C_\theta(\Delta,T) \, , \tag{B3}$$

where $C_\theta(\Delta,T) = D\partial G_3/\partial Z|_{Z=0}$ is the rate ( i.e., the probability per unit diffusion time step) that a particle diffusing in one dimension that started at $z=\Delta$ at $t=0$ is absorbed at the origin $z=0$ at a later time $T > 0$. This probability is calculated in a straightforward manner by making use of the independence of the successive returns of a random walk to its starting point, as has been done in Sec. 3.1.

The total probability of re-adsorption of a ligand, averaged over all space, is thus

$$\gamma(t) = k_{-} \int_0^t d\tau p(\tau) \int dx dy \tilde{P}(x,y;t-\tau) \quad . \tag{B4}$$

After substituting Eq. B3 into Eq. B4 and performing the trivial spatial averaging, we finally arrive at Eq. 2.

## Appendix C: Density profile of ligands in solution

In this appendix, we show how the density profile of the ligands in the z-direction could be computed within our formalism for arbitrary times. The general expression for the density $\rho(z,t)$ of ligands at a distance z above the surface at time t has the form

$$\rho(z,t) = \frac{k_{-}\theta_s}{\Delta^2} \int_0^t d\tau p(\tau) F_\theta(z, t-\tau) \quad , \tag{C1}$$

where $F_\theta(z,T)$ is the Greens function for diffusive transport of a ligand following dissociation at the surface at time $t=0$ to the height $z$ at time $t=T$, in the presence of the partially absorbing boundary at z=0. This function is expressed via the self-consistent equation

$$F_\theta(z,t) = F^0(z,t) + (1-\theta) \int_0^t d\tau q(\Delta, \tau) F_\theta(z, t-\tau) \quad , \tag{C2}$$

where

$$F^0(z,t) = (4\pi Dt)^{-1} \left[ e^{-\frac{(z-\Delta)^2}{4Dt}} - e^{-\frac{(z+\Delta)^2}{4Dt}} \right] \tag{C3}$$

is the probability of the ligand reaching z at time t, without ever touching the surface in between. Eq.(C1) and Eq.(C2) may be solved together using the Laplace transform method, as explained in the main text. The solution in terms of Laplace-transformed variables is

$$\tilde{\rho}(z,s) = \frac{k_{-}\theta_s}{\Delta^2} \frac{\tilde{F}^0(z,s)\tilde{p}(s)}{1-(1-\theta)\tilde{q}(s)} \quad . \tag{C4}$$

From Eq.C3, we find that

$$\tilde{F}^0(z,s) = \frac{e^{-z\sqrt{s/D}}}{\sqrt{Ds}} \sinh\left(\Delta \sqrt{\frac{s}{D}}\right) \approx \frac{\Delta}{D} e^{-z\sqrt{s/D}} \quad , \tag{C5}$$

where the last step is valid for sufficiently late times $t \gg \Delta^2/D$, which we assume to be satisfied. In order to check consistency with the previous discussions, let us compute the Laplace transformed concentration close to the substrate surface $\tilde{\rho}(\Delta,s)$, which is given by $\tilde{\rho}(\Delta,s) = k_{-} \frac{\theta_s}{D\Delta} \tilde{p}(s) \frac{e^{-2\sqrt{\delta s}}}{1-(1-\theta)\tilde{q}(s)} = K_D \tilde{p}(s) \tilde{C}_\theta(s)$. Upon inversion, this relation gives $\rho(\Delta,t) = K_D \int_0^t d\tau p(\tau) C_\theta(t-\tau)$, which is consistent with our argument in Sec.3.1, following Eq.2.

We now combine Eq. C4 and Eq. C5 with Eq. 7 and Eq. 10. In the regime of non-microscopic time as mentioned above, the leading term is

$$\tilde{\rho}(z,s) \approx K_D \frac{p(0)}{s+\sqrt{cs}} e^{-z\sqrt{s/D}} + O(\sqrt{\delta c}) , \qquad (C6)$$

where the constant c is given by Eq.11. The inversion of this expression (after leaving out the $O(\sqrt{\delta c})$ term) gives

$$\rho(z,t) \approx p(0) K_D e^{\frac{\sqrt{c}}{D}z^2 + ct} \, \text{erfc}\left[\sqrt{ct} + \frac{z^2}{2\sqrt{Dt}}\right] , \qquad (C7)$$

which is the density profile of ligands in the bulk solution. At late times $t \gg c^{-1}$, we obtain the asymptotic form

$$\rho(z, t \gg c^{-1}) \approx p(0) \frac{\theta_s}{\Delta^2} \frac{e^{-\frac{z^2}{4Dt}}}{\sqrt{\pi Dt}} . \qquad (C8)$$

## Appendix D: Exponential versus non-exponential decay

In this appendix, we discuss in detail the different time regimes of decay of the bound fraction. We consider non-microscopic times $t \gg \delta$ (corresponding to $\delta s \ll 1$) so that the expansion $e^{-2\sqrt{\delta s}} = 1 - 2\sqrt{\delta s}$ may still be used in Eq.9. Up to $O(\sqrt{\delta s})$, we then have

$$1 - \tilde{C}_\theta(s) \approx \frac{2\sqrt{\delta s}}{\theta + 2(1-\theta)\sqrt{\delta s}} . \qquad (D1)$$

When $\theta$ is sufficiently large, clearly the first term in the denominator dominates over the second. In this limit, we thus recover the form in Eq.10.

However, when $\theta \ll 1$, the second term dominates for sufficiently large $s$, i.e., for $s \gg \theta^2/4\delta$ (which corresponds to times $t \ll 4\delta/\theta^2 \equiv t_e$). In this limit, $1 - \tilde{C}_\theta(s) \approx 1 + O\left(\frac{\theta}{\sqrt{\delta s}}\right)$, which, after substitution in Eq.4 gives exponential decay. In the opposite limit of sufficiently small $s$, however, the first term dominates (over times $t \gg t_e$), so that we recover the forms in Eq.10 and Eq.11.

To summarize, therefore, the different regimes of decay for $\theta \ll 1$ are

$$p(t) \approx p(0)e^{-k_- t} \qquad\qquad t \ll t_e \equiv 4\delta/\theta^2 \ , \qquad\qquad \text{(D2a)}$$
$$p(t) = p(0)e^{ct}\,\text{erfc}(\sqrt{ct}) \qquad\qquad t \gg t_e \ , \qquad\qquad \text{(D2b)}$$

where the constant c is defined in Eq.11.

We conclude that exponential decay of the bound fraction with the intrinsic dissociation rate may be recovered in the limit of sufficiently small surface coverage, at sufficiently small times. For any non-zero surface coverage, however, the long-time decay always has the non-exponential form in Eq.11.

Another possibility whereby one may recover exponential decay of the bound fraction is to start with an initial bound fraction p(0) ~ 1, so that very few sites are initially available for rebinding. In this case, the initial part of the dissociation curve may be expected to follow the purely exponential dissociation, with the intrinsic rate. However, this method does not always work in practice for two reasons: To reach a steady state with a high value of p(0), one needs to use a large (often impractically high) concentration of ligands, and the steady state itself may then become difficult to reach over a reasonable interval of time. Secondly, even if such a high initial p(0) could be attained, a reliable measurement of the dissociation rate would require the observation of the dissociation curve over a time scale $\sim k_-^{-1}$, by which time a fraction 1/e~ 37% of the binding sites have released the ligands and rebinding is already significant, unless the association rate and the surface coverage are sufficiently small. Indeed, the inadequateness of using only a part of the dissociation curve for data fitting has been pointed out by other authors as well [van der Merwe 2000].

**Tables:**

| Quantity | Symbol | typical units |
|---|---|---|
| Microscopic length scale | $\Delta$ | m |
| Diffusion coefficient | D | $m^2 s^{-1}$ |
| Microscopic time scale | $\delta = \Delta^2/4D$ | s |
| Association rate | $k_+$ | $M^{-1} min^{-1}$ |
| Dissociation rate | $k_-$ | $min^{-1}$ |
| Equilibrium dissociation constant | $K_D = \dfrac{k_-}{k_+}$ | M |
| Fractional surface coverage | $\theta_s$ | dimensionless |
| Effective surface coverage (model parameter) | $\theta = \theta_s \dfrac{k_+}{D\Delta}$ | dimensionless |
| Bound receptor fraction at time t | $p(t)$ | dimensionless |
| Ligand density profile at time t and height z above the surface (z = 0) | $\rho(z,t)$ | Number of molecules/$m^3$ |
| Rebinding rate | $\gamma(t) = k_+ \rho(\Delta,t)$ | $s^{-1}$ |
| Adsorption rate | $C_\theta(\Delta,t)$ | $s^{-1}$ |
| Dimensionless dissociation rate | $\tilde{\beta} = k_- \dfrac{\Delta^2}{6D}$ | dimensionless |

**TABLE 1**: A glossary of the important quantities discussed in the paper, along with the corresponding units.

| Heparin level (μM) | Fit parameter $c_H$ ($s^{-1}$) | $\theta_H/\theta_0 = \sqrt{c_0/c_H}$ |
|---|---|---|
| 0.0 | $1.9 \times 10^{-5} = c_0$ | 1.0000 |
| 1.8 | $2.9 \times 10^{-4}$ | 0.2559 |
| 3.6 | $4.1 \times 10^{-4}$ | 0.2140 |
| 5.4 | $5.9 \times 10^{-4}$ | 0.1934 |
| 10.8 | $8.0 \times 10^{-4}$ | 0.1495 |

**TABLE 2**: Fit parameters to SPR experimental data for various heparin concentrations. Note that the 'effective coverage' decreases with the heparin concentration (since the ligand bound to heparin is unavailable for binding to surface proteins), but the drop is much less rapid than a prediction based on complete equilibration between the heparin and IGF concentrations would suggest.

**Figure legends:**

FIG. 1: (a) A schematic diagram of the SPR experimental set-up showing the attached ligands (IGF-I) available for binding to the IGFBP-3 in solution. (b) Representative association-dissociation plot of IGFBP-3 (20 nM) for surface-coupled IGF-I under flow conditions (0.75 ml/min). The arrow labeled PBST indicates when the fluid was changed from IGFBP-3 in PBST to only PBST, thus initiating the dissociation phase.

FIG. 2. (a). Representative plot of dissociation phase data for IGFBP-3 under flow (0.75 ml/min) and non-flow conditions, both normalized to peak value. (b) Representative plot of association phase data for IGFBP-3 (50 nM) +/- heparin (200 nM) or heparin alone to IGF-I (off-line coupling) under flow conditions (0.75 ml/min).

FIG. 3. Representative data for dissociation phase of IGFBP-3 from IGF-I for PBST (buffer alone) or heparin (30 μM) in PBST for duplicate runs of each on the same chip normalized to the individual time 0 value in the absence of flow. The data is representative of multiple runs performed on six independent chips.

FIG. 4. Comparison of dissociation data in the presence of heparin (30 μM) for two different levels of surface coupled IGF-I (on-line coupling): (◊) ~ 4 pixels of surface coverage and (+) ~12 pixels of surface coverage. This observation is consistent with the mean-field calculation in Sec. 3.1 in the text. Results for other heparin concentrations, as well as runs without heparin, showed similar trends.

FIG. 5. A schematic diagram illustrating the setup of our mean-field model. The receptors are modeled as point size absorbing objects on the substrate surface, and the ligands diffuse in the bulk solution above this surface.

FIG. 6. Normalized dissociation curve from simulations of the lattice model for two different surface protein densities. The initial bound fraction p(0) is 0.25 in both cases, and the dissociation rate is $\tilde{\beta} = 10^{-5}$. The thin lines indicate optimal fits using Eq. 11, with c = 0.01 for θ = 0.1 and c = 0.0004 for θ = 0.5. The corresponding theoretical values are c = 0.006 and c = 0.00024, respectively. The results represent averages over 20 different starting configurations.

FIG. 7. The effect of reducing the dissociation coefficient relative to the surface coverage (which is fixed at 1% here) in the simulations. We observe that when the dissociation rate is high, the temporal decay becomes effectively exponential (compare with the dashed exponential curve) in accordance with the mean-field calculations in Appendix B. As $k_-$ is reduced, rebinding is increasingly important, and the dissociation slows down. The results were averaged over 20 different starting configurations.

FIG. 8. Simulation dissociation curves for two small coverage fractions, 0.5% and 1% of the surface area. The dissociation rate is $\tilde{\beta} = 10^{-5}$.

FIG. 9. The same data as in Fig. 8 is plotted on a logarithmic scale. This plot shows the cross-over to the power-law regime mentioned in Sec. 3.1. The straight line is a fit function $f(T) = T^{-1/2}$.

FIG.10. Comparison of mean-field theory with experimental SPR data (◊) for IGFBP-3 dissociation from IGF-I in the absence of heparin. The thin curve represents the best fit using Eq. 11, with c = $1.9 \times 10^{-5}$ s$^{-1}$, and the straight line is the best exponential fit, with a dissociation rate $k_- \approx 0.066\, min^{-1}$. The experimental data was averaged over two different runs on the same IGF-I coupled chip and is representative of averaged data from six separate chips.

FIG.11. Comparison of mean-field theory with IGFBP-3 dissociation SPR data in the presence and absence of heparin (concentration of heparin indicated on figure by experimental values; the topmost curve is a reproduction of the zero heparin data in Fig.10). The lines represent the fitting curves using Eq.(11), with fit parameters c= $4.1 \times 10^{-4}$ s$^{-1}$ and c=$5.9 \times 10^{-4}$ s$^{-1}$ respectively for 3.6 μM and 5.4 μM heparin.

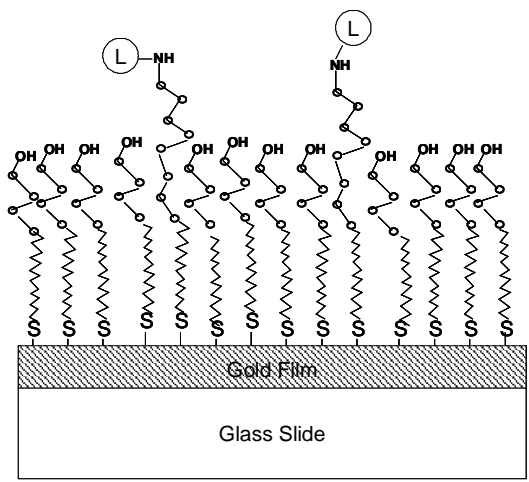

**(A)**

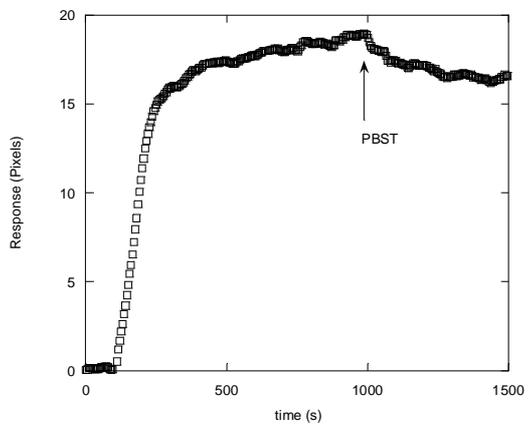

**(B)**

**FIG1.**

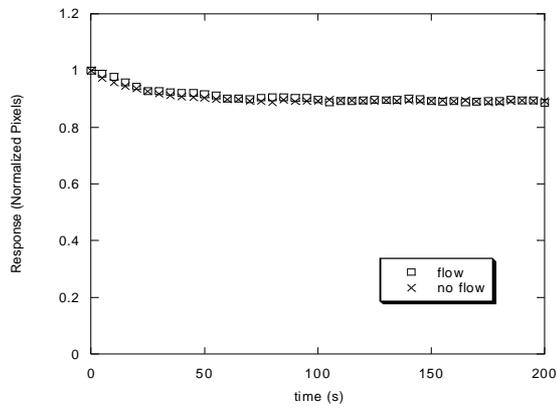

**(A)**

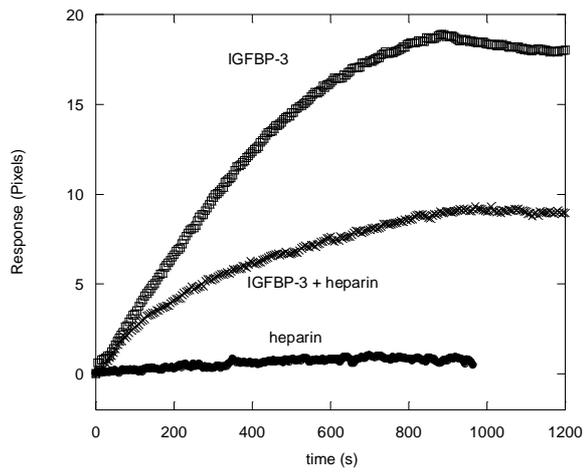

**(B)**

**FIG2.**

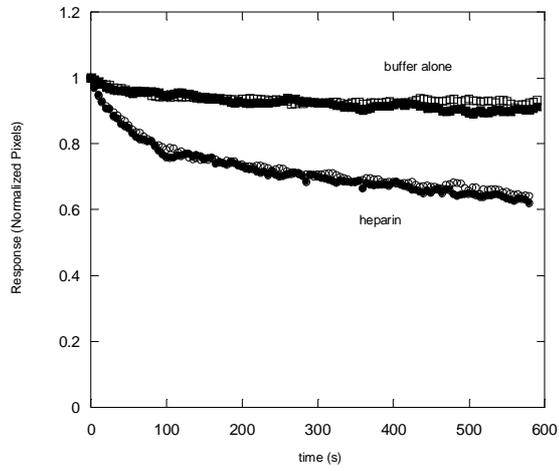

**FIG3.**

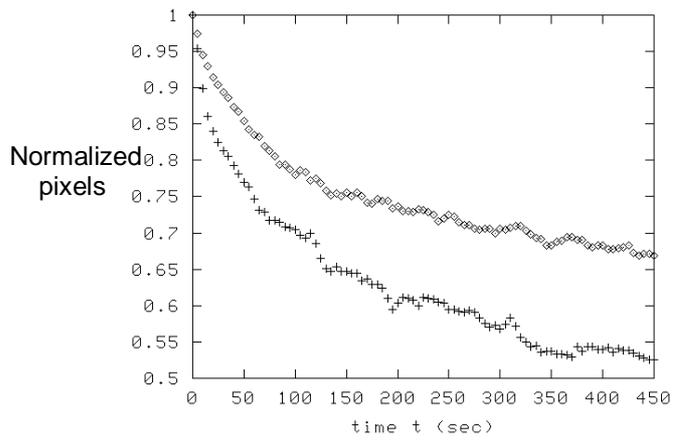

**FIG4.**

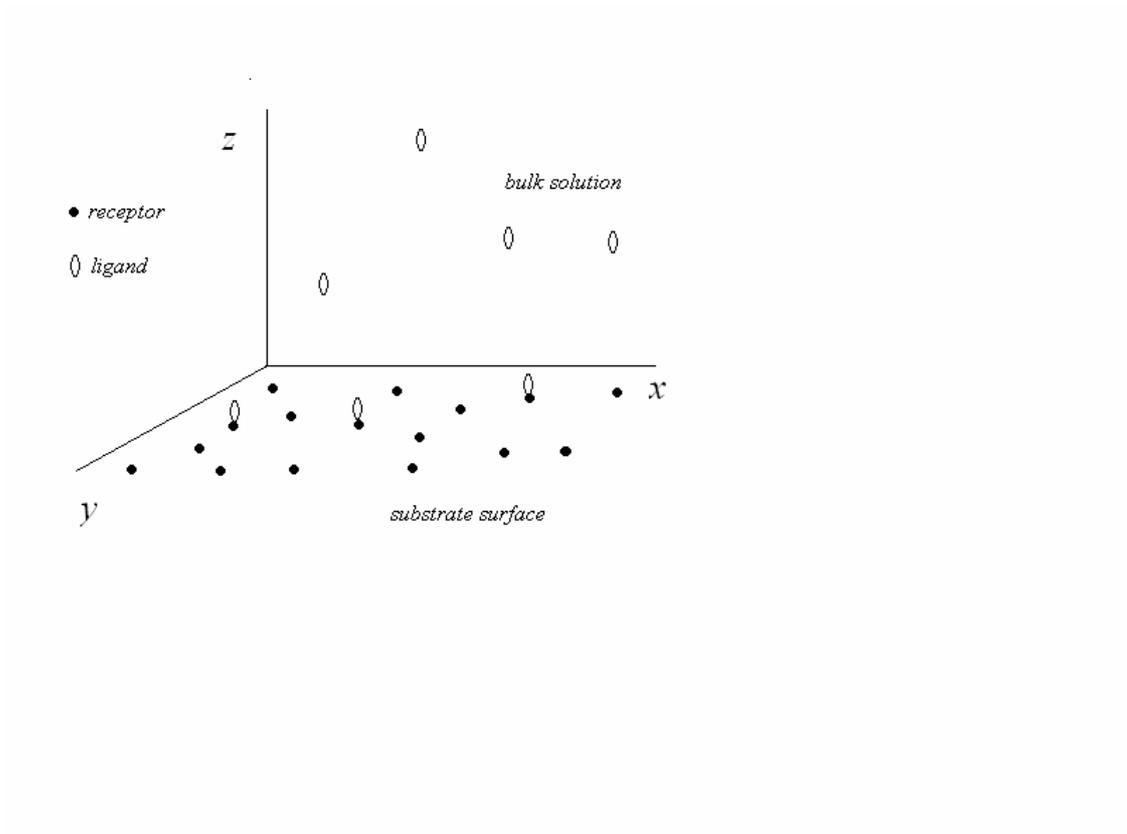

**FIG5.**

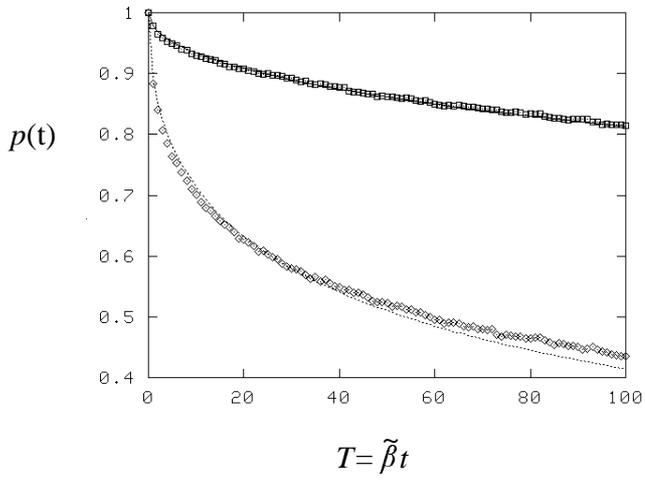

**FIG6.**

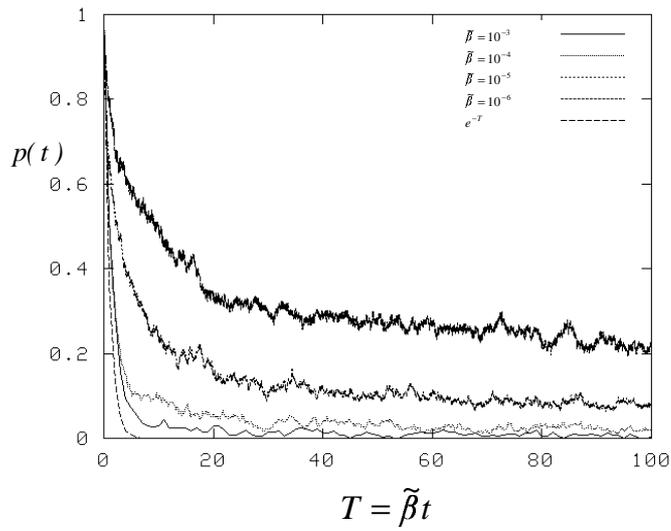

**FIG7**.

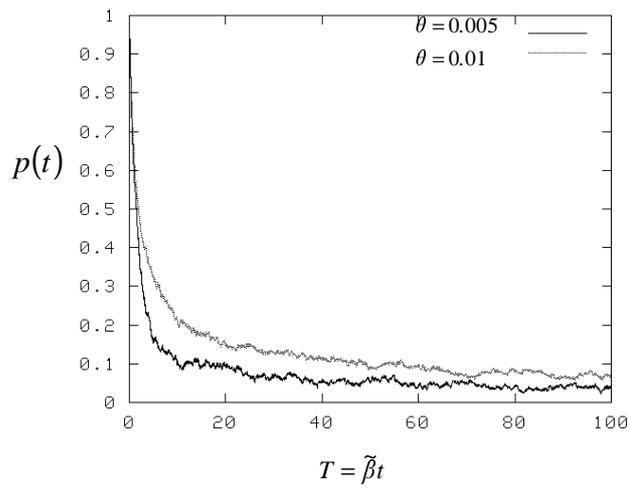

**FIG8**.

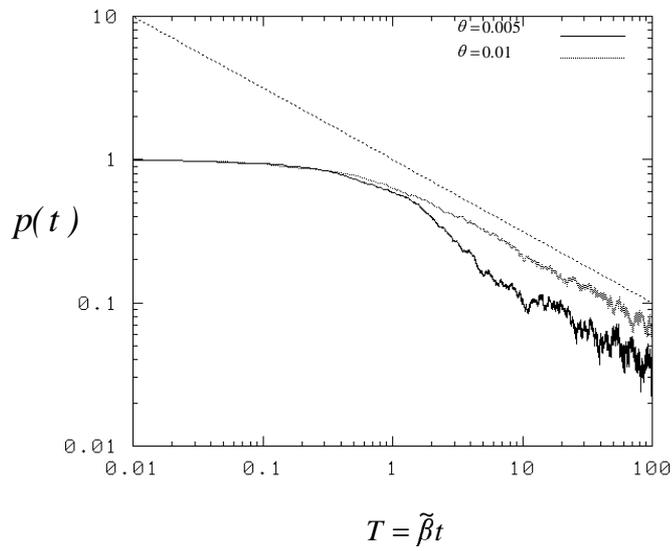

**FIG9**.

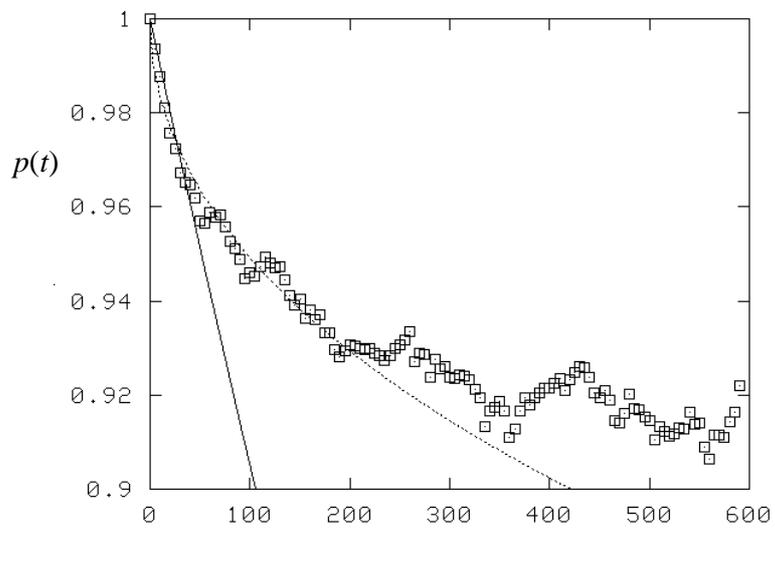

**FIG10**.

$t$(sec)

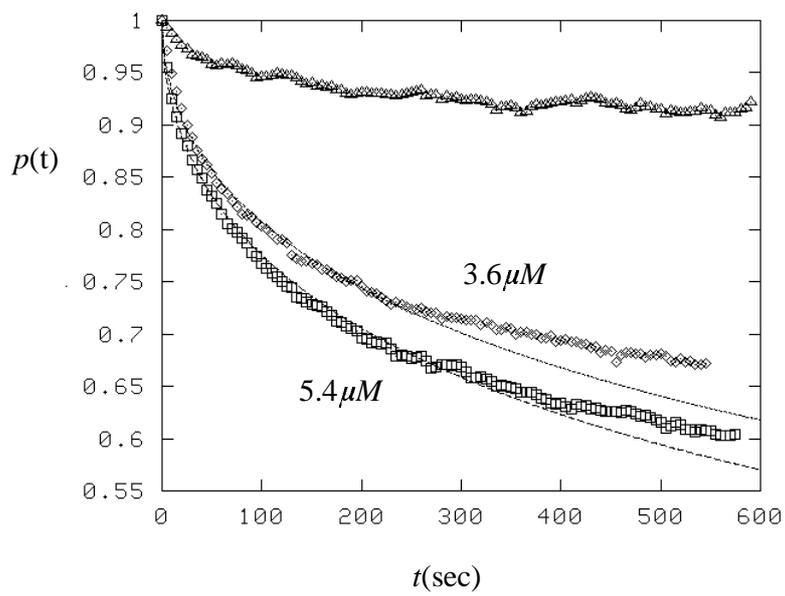

**FIG11.**